\input amstex
\documentstyle{amsppt}

\redefine\w{\omega}
\define\nn{\medpagebreak}
\define\n{\smallpagebreak}
\define\nnn{\bigpagebreak}
\define\a{\alpha}
\define\g{\gamma}
\define\ar{\longrightarrow}

\redefine\b{\beta}
\redefine\P{\operatorname{P}}
\define\NP{\operatorname{NP}}

\define\PLSN{\operatorname{PLSN}}
\define\PLS{\operatorname{PLS}}
\define\NPLSN{\operatorname{NPLSN}}
\define\NPLS{\operatorname{NPLS}}
\define\PP{\operatorname{\Pi P}}
\define\SP{\operatorname{\Sigma P}}
\define\PH{\operatorname{PH}}
\redefine\Im{\operatorname{Im}}
\redefine\max{\operatornamewithlimits{max}}
\define\SAT{\operatorname{SAT}}

\document
\topmatter
\title
Lindenmayer systems as a model of computations
\endtitle
\author
Y. Ozhigov
\endauthor
\abstract

LS is a particular type of computational processes simulating living 
tissue. They use an unlimited branching process arising from the 
simultaneous substitutions of some words instead of letters in some 
initial word. This combines the properties of cellular automata and 
grammars. It is proved that

1) The set of languages, computed in a polynomial time on such LS that all 
replacing words are not empty, is exactly NP- languages.

2) The set of languages, computed in a polynomial time on arbitrary LS ,
contains the polynomial hierarchy.

3) The set of languages, computed in a polynomial time on a nondeterministic 
version of LS, strictly contains the set of languages, computed in a 
polynomial time on Turing Machines with a space complexity $n^a$, 
where $a$ is positive integer.

In particular, the last two results mean that Lindenmayer systems
may be even more powerful tool of computations than nondeterministic
Turing Machine.

\endabstract
\address
Ozhigov Yuri., Department of Mathematics, Moscow State
                Technological University "Stankin", Vadkovsky per. 3a,
                 101472, Moscow, Russia
\endaddress
\email
y\@ oz.msk.ru
\endemail
\endtopmatter

\specialhead
\ \ \ \ \ \ \ \ \ \ \ \ 1.Introduction and definitions\endspecialhead
\nnn

The simulation of nondeterminism by a physical device remains attractive but 
unattainable dream. Even the most powerful physical computational device - 
quantum computer is not able to achieve this objective in the computations 
with oracles as follows from the result of C.Bennett, E.Bernstein,
G.Brassard and U.Vazirani \cite{1}.
This is most likely to be true also for absolute (without oracles) 
computations. Therefore living things present, probably the best real 
model of nondeterminism in nature. 

The simple mathematical model of the structures with
two fundamental properties of a Life: reproductions and determined 
stable interactions with the nearest neighborhood, was proposed by 
A.Lindenmayer (\cite{4},\cite{5}\cite{6} ). 
One-dimensional Lindenmayer systems considered in this paper is a crude
model of a reproduction because we ignore the spatial arrangement of 
tissue,
otherwise this model is well suited to describe the behavior of living 
cells. 
 The spatial arrangement of tissue can be often ignored for example,
for the filamentous organisms.

Lindenmayer system also bears a resemblance to an imaginary nondeterministic
 computer with the supernatural abilities. These abilities is the subject 
for study in this
work. We proceed with an exact definition.

 Let $\w =\{ a_0 ,a_1 ,\dots ,a_n \}$ be our basic alphabet, $\w ^*$ denotes
the set of all words from $\w$ and $\Sigma =\{ a_0 Aa_0 \ |\ A\in\w ^* \}$ 
will be our
basic set of words. If $x\in\w ^*$, then $|x|$ means the length of the word 
$x$.
Let $\Cal A$ be a function of the form:
$$
\Cal A :\ \w ^3 \ar\w ^* ,
$$
where $\Cal A (a_i ,a_0 ,a_j )=a_0$ for all $i,j=0,1,\dots ,n$. Then $\Cal A$ 
is called
LS. If here $\forall x\in\Im\Cal A \ |x|>0$ then 
$\Cal A$ is called L-system with nonempty results - LSN.

Finally, if here $\forall x\in\Im\Cal A\ |x|=1$, then $\Cal A$ is an ordinary 
cellular automaton (CA) (look at \cite{10} ).

Evolutions generated by LS ( L-languages ) and their algorithmic properties  
were subject for study mainly since 1971.
 As a typical example of mathematical problem
here we refer to an algorithmically undecidable problem of coincidence for 
languages, 
determined by two LS ( look at the work of A.Lindenmayer \cite{6}). 
W.J.Savitch in \cite{8}  gave some conditions which should be imposed on the 
nontrivial set of strings to ensure that it is L-language; this conditions use 
stack machines. J. van Leeuwen in \cite{3} obtained upper bounds for space 
required to recognize some extensions of L-languages  ( by natural operations 
such as sum and intersection ) deterministically and nondeterministically. 
P.Vitanyi in \cite{9} treated context sensitive table L-languages. Other 
interesting examples may be found in \cite{7}.

In the present paper LS will be considered as a particular type of 
computations using unlimited branching processes which arise if  the 
length of some word $\Cal A (a_i a_j a_k )$ is more than 1. Such a process 
requires exponentional space but reduces the time for solutions of NP-complete 
problems. The advantage of such computational processes are their simplicity 
and power. The simplicity of rules for LS allows to find them by evolutionary 
programming as in the work \cite{2} of J.R.Koza.
\n

In this paper we study a polynomial complexity of computations on LS.
LS may be considered as a sort of cellular automata with reproductions  which 
resembles a nondeterminism.
It is founded that the set of languages computed in a polynomial time on
deterministic as well as on nondeterministic version of LSN is coincident with 
NP.
Thus the complexity status of LSN is clear with a precision of 
$\P\overset ?\to =\NP$-problem.
The following peculiarities of LS must be mentioned explicitly.

The first is that the possibility of an empty results 
$\Cal A (a_i ,a_j ,a_k )$
 means the annihilation of some domains
in a computational space and so it allows to transmit an information faster 
comparatively
with LSN. In the section 4 it is proved that the set of languages computed 
on LS in a polynomial time contains the polynomial hierarchy.

The second is that this presumably makes LS inconvenient
for analysis and its complexity status is still obscure.

Let $\P _\a$ be the set of languages, which can be computed on Turing Machines 
in a polynomial time with a space complexity $n^\a ,\ \a =1,2,\dots .$
The sole lower bound of complexity result having to do with LS is that the set 
of languages computed
on the nondeterministic version of LS in polynomial time strictly includes 
$\P _\a$.
\nn

\proclaim{Remark} All inclusions established in this paper can be relativized. 
This property 
may be traced by a reader in all constructions.
\endproclaim
\n

If $\Cal A$ is LS and
$$
B=a_{i_0} a_{i_1} a_{i_2} \dots a_{i_p} a_{i_{p+1}}
\tag 1
$$
is an arbitrary word from $\Sigma$, $a_{i_0} =a_{i_{p+1}} =a_0$, the new 
word $\Cal A (B)$, defined by
$\Cal A (B)=a_0 A_1 A_2 \dots A_p a_0
$
where $\forall j=1,\dots p\ \ A_j =\Cal A (a_{i_{j-1}} ,a_{i_j} ,
a_{i_{j+1}} )$,
is called the result of $\Cal A$-action on $B$.

Let $W$ be the underlined occurrence in the word $B$:
$$
a_0 a_{i_1} \dots a_{i_{s-1}} \underline{ a_{i_s} a_{i_{s+1}} 
\dots a_{i_t} }
a_{i_{t+1}} \dots a_{i_p} a_0 .
$$
The base of $W$ is the word $a_{i_s} a_{i_{s+1}} \dots a_{i_t}$. Then 
the underlined occurrence $W_1$ in the word $\Cal A (B)$:
$$
a_0 A_{i_1} \dots A_{i_{s-1}} \underline{ A_{i_s} A_{i_{s+1}} \dots A_{i_t} }
A_{i_{t+1}} \dots A_{i_p} a_0
$$
is called the descendant of $W$ in transformation $B\ar\Cal A (B)$.
If $A_j \neq a_{i_j}$ then an occurrence of $a_{i_j}$ in $B$ will be 
considered
to be affected in this transformation.

A sequence 
$$
B_0 ,B_1 ,\dots ,B_N
$$
is called an evolution of $\Cal A$ if 
$\forall i=1,\dots ,N\ \ B_i =\Cal A (B_{i-1} )$.
If $B=B_0$, the descendant of $W$ in this evolution is defined by induction on 
$N$.
This evolution is called complete if the letter $a_n$ occurs only in the word 
$B_N$ and not
in $B_1 ,\dots ,B_{N-1}$.

We say that a set $M$ of some words in alphabet 
$\w _1 =\{ a_1 ,\dots ,a_{n-1} \}$ is 
computed on LS $\Cal A$ if for any $B\in\w _1^*$ the following conditions
are equivalent:

1) there exists a complete evolution $B_0 ,B_1 ,\dots ,B_N$, where $a_0 Ba_0 
=B_0$.

2) $B\in M$.

In this case for $B\in M$ we denote the number $N$ from 1) by 
$\tau _\Cal A (B)$.
The set $M$, computed on $\Cal A$, is denoted by $M_\Cal A$. 

The time complexity of LS $\Cal A$ is defined by the following
$$
t_\Cal A (n)=\max\limits _{|B|\leq n,\ B\in M} \tau _\Cal A (B) .
$$

Let PLS ( PLSN ) denote the class of languages, computed on LS ( on LSN )
in a polynomial time.
\nn

Now let us define all corresponding notions for a nondeterministic version of 
LS ( LSN ).
To denote these notions we add the letter "N" : NLS, NLSN, etc. 

For an arbitrary set $C$ let $2^C$ denote the set of all subsets of $C$.
Thus, NLS is 
a function of the form
$$
\Cal A :\ \w ^3 \ar 2^{\w ^*} .
$$

If $\forall \bar b \in\w ^3 \ \ \forall C\in\Cal A (\bar b ) \ \ |C|>0$, 
then
$\Cal A $ is called NLSN.
If a word $B$ from $\Sigma$ has the form (1) then the result of 
$\Cal A$-action on $B$ is the set 
$$
\Cal A (B) =\{ a_0 A_1 A_2 \dots A_p a_0 \ |\ \forall j=1,\dots ,
p\ A_j \in\Cal A (a_{i_{j-1}} ,a_{i_j} ,a_{i_{j+1}} ) \} .
$$

A sequence $B_0 ,\dots B_N$ is called an evolution of $\Cal A$ if 
$\forall i=1,\dots ,N \ \ B_i \in\Cal A (B_{i-1} )$. A complete evolution 
is defined as above.

 If $B\in\w ^*$ let $\tau _\Cal A (B)$ be the minimal length of complete 
evolutions
beginning with $B_0 =a_0 Ba_0$ if such an evolution exists, and 
$\tau _\Cal A (B) =\infty$ in
the opposite case.

NLS $\Cal A$ admits a word $B$ iff $\tau _\Cal A (B) <\infty$. Thereafter, 
the time complexity of $\Cal A$ can be defined by
$$
t_\Cal A (n)=\max\{ \tau _\Cal A (B)\ |\ |B|\leq n,\ 
\tau _\Cal A (B)<\infty\} .
$$
$M_\Cal A $ denotes here the set of all words admitted by $\Cal A$ , we say
that this set is computed by $\Cal A$.

So, NPLS(N) denotes the class of all sets, computed on NLS(N) in a polynomial 
time.
The first two results to be established in this paper are the following.
\nn

\proclaim{Theorem 1} $\PLSN =\NPLSN =\NP$ \endproclaim
\n

\proclaim{Theorem 2} $\forall \a =1,2,\dots\ \ \NPLS\neq\P _\a$ \endproclaim
\nnn

\specialhead
\ \ \ \ \ \ \ \ \ \ \ \ 2. Deterministic version of LSN \endspecialhead
\nn

\head Proof of Theorem 1 \endhead
\nn

The deduction of inclusion
$$
\NP \subseteq\PLSN
\tag 2$$
will be our initial concern. Note that the establishment of the belonging
$ c\in\PLSN$, where $C$ is an arbitrary NP-complete set would be ample.
For example, let SAT be the set of all satisfiable formulas of a 
propositional
logic. The inclusion (2) reduces to the establishment of the following
belonging:
$$
\SAT\in\PLSN .
$$

Let $\bar l =\{ \& ,\vee ,\neg , (, ) ,0,1\}$ be the basic alphabet for a 
propositional
logic. Let any Boolean variable $\a _i$ be coded by the word
$0\undersetbrace i\ \text{times}\to{11\dots 1} 0$. Then any Boolean formula
$\phi$ can be encoded in $\bar l$ and let $\ulcorner\phi\urcorner$ be its
code. Let $\bar s =\{ f,t\}$ be an auxiliary alphabet for Boolean values:
$t$ - true and $f$ - false.
\nn

Let a special alphabet $\w _1$ consist of the following parts:

1) the alphabet $\bar l$ for the propositional logic;

2) the alphabet $\bar s$ for the Boolean values;

3) auxiliary letters $\a ,\b ,b, e,h,*,+,\% ,\nu $;

4) alphabets of doubles: $L_1 ,L_2 ,L_3 ,L_4 ,L_5$ for the letters from

$L=\bar l \cup\{\a ,h,b,e,\nu\}\cup\bar s$
(if $x\in L$ we'll denote its doubles by $x' \in L_1 ,\tilde x \in L_2 ,x^0 
\in L_3 ,x^+ \in L_4 ,\bar x\in L_5$ respectively ).

5) the alphabet $L_5 \times L_5$ whose elements are denoted by 
$\binom{d_1}{d_2}$ , where 
$d_1 ,d_2 \in L_5$.

\proclaim{Lemma 1} There exists LSN $\Cal A _1$ with the alphabet $\w _1$, 
such that for any propositional formula $\phi$ ( which has been encoded as 
it is pointed out above ) there exists the following evolution of 
$\Cal A _1$ :
$$
B_0 = a_0 \ulcorner\phi\urcorner a_0 , B_1 ,\dots , B_N ,\dots
$$
where:

1) $N=4|\ulcorner\phi\urcorner |^2 +7|\ulcorner\phi\urcorner |+15$,
$B_N$ has the form 
$$
a_0 e A_1 eA_2 e\dots eA_q ea_0 ,
\tag 3$$
$q=2^{|\ulcorner\phi\urcorner |} ,\ \forall i=1,2,\dots ,q 
\ A_i =S_i \% \ulcorner\phi\urcorner ,
\ S_i \in\bar s ^* ,\ |S_i |=|\ulcorner\phi\urcorner |$, and for all 
$i,j$ if $1\leq i<j\leq q$ then $S_i \neq S_j$.

2) $\forall j=2,3,\dots ,N,\ A,B$ if 
$$
B_j =A\% B
\tag 4$$
 then $\exists A',B' :\ A=A'S' , \ B=\ulcorner\phi\urcorner B'$,
where $S' \in\bar s ^* ,\ |S' |=|\ulcorner\phi\urcorner |$, and the 
underlined
occurrence in the word 
$B_j :\ A' \underline{ S' \% \ulcorner\phi\urcorner } B'$ 
is not affected in the evolution $B_j ,B_{j+1} ,\dots $.

3) $\forall j>N \ B_j =B_N .$
\endproclaim
\nn

\head Proof of Lemma 1 \endhead
\nn

It is not difficult to understand how such an automaton $\Cal A _1$ can be 
constructed.
It must replicate the words of the form $Sb^k \a\phi$ for various values of 
$S\in\bar s ^*$ sequentially, where for all time 
$|Sb^k |=|\ulcorner\phi\urcorner |, \ |S|=1,2,\dots ,\ b$ is 
the special letter, $k\in\Bbb N$. And what is essential for the present 
purposes, this can be done in a polynomial time for the input data 
$\ulcorner\phi\urcorner$ by the definition of LSN.

Any LS $\Cal A$ with an alphabet $\w$ may be determined by the list of all 
records of the form
$$
(a,b,c)\ar\Cal A (a,b,c);\ a,b,c\in\w ,\Cal A (a,b,c)\neq b,
$$
which may be thought of as active commands. In what follows any list of 
commands is taken to be
supplemented by all plausible (passive) commands of the form
$$
(a,b,c)\ar b 
$$
which are not written explicitly.

The list of commands, determining $\Cal A _1$ consists of two groups:

\subsubhead Group 1 \endsubsubhead

\n
{\it input: $a_0 \ulcorner\phi\urcorner a_0$,}

{\it output: $a_0 \undersetbrace {|\ulcorner\phi\urcorner |}\to{b\dots b} \a 
\ulcorner\phi\urcorner ea_0$}

{\it time: $\leq 7|\ulcorner\phi\urcorner |+5$.}

$x,y$ take all values from $\bar l$, $z$ takes all values from $\bar l$ and 
its doubles from $L_2$,
? - from $\w _1$.
\n

$$
\aligned
(x,a_0 ,?) &\ar ea_0 ,\\ (?,a_0 ,x) &\ar a_0 + ,\\
(?,+,z) &\ar\b +, \\ (?,a_0,\b ) &\ar a_0 e ,\\ (+,x,?) &\ar x' ,\\
(x' ,y,?) &\ar\tilde y ,\\ (?,\tilde x,?) &\ar x' ,\endaligned\ \ 
\aligned
(?,x',\tilde y) &\ar \tilde x ,\\ (?,x',e) &\ar x^0 ,\\
(?,x^0 ,?) &\ar x ,\\ (?, x' ,y^0 ) &\ar x^0 ,\\
(?,+,x^0 ) &\ar \a ,\\ (?,\b ,\a ) &\ar b ,\\ (?,\b ,b) &\ar b .
\endaligned
$$

\nn

\subsubhead Group 2 \endsubsubhead
\n

{\it input: } $a_0 eb^q \a \ulcorner\phi\urcorner ea_0$,

{\it output: } (3)

{\it time: } $\leq 4|\ulcorner\phi\urcorner |^2 +10$,

$w,x,y,z,u$ take all values from $L-\{ e\}$.
\n

$$
\aligned
(v,b,?) &\ar \tilde \nu \ (v\in\w _1 -\{b,\b\} ),\\
(?,x,\tilde y ) &\ar \tilde x ,\\
(r,\tilde b ,?) &\ar \tilde \nu ,\ (r\in\bar s\cup\{ e,*,a_0 \} ) ,\\
(\rho ,\tilde x,?) &\ar \bar x ,\ \rho\in\{ *,e\} ,\\
(\bar x ,y,?) &\ar \bar y ,\\
(\bar x ,\bar y ,e) &\ar \binom{\bar y}{\bar y} \ (x\notin\bar s\ \text{or}\ 
y\neq b ),\\
(\bar x ,e,?) &\ar ** ,\\
(\binom{\bar x}{\bar y} ,*,?) &\ar *y ,\ (y\neq \nu ),\\
(\binom{\bar x}{\bar\nu} ,*,?) &\ar *t ,\\
(\bar w ,\bar x,\binom{\bar y}{\bar z} ) &\ar\binom{\bar x}{\bar h} ,
\endaligned \ \ \aligned
(x,\binom{\bar y}{\bar z} ,?) &\ar\binom{\bar y}{\bar x} ,\\
(\binom{\bar x}{\bar d} ,\binom{\bar y}{\bar g} ,?)
 &\ar\binom{\bar y}{\bar d} ,\ d,g\in L\cup\{ h\} ,\\
(c,\bar x ,\binom{\bar y}{\bar h} ) &\ar x ,\\
 x\neq \nu &,\ c\in\{ e,*,a_0 \} ,\\
(c,\bar\nu ,\binom{\bar x}{\bar h} ) &\ar f ,\ c\in\{ e,*,a_0 \} ,\\
(x,\binom{\bar y}{\bar z} ,?) &\ar y ,\ x\notin\bar s \ \text{or}\ y\neq \nu ,\\
(r,\binom{\bar \nu}{\bar x} ,?) &\ar f ,\ r\in\bar s ,\\
(x,*,?) &\ar e ,\\
(\bar r ,\bar b ,e) &\ar \binom{\bar b}{\bar \nu } ,\\
(x,\a ,?) &\ar \% .
\endaligned
$$
\nn

\subsubhead Informal comments \endsubsubhead

Let an evolution of $\Cal A _1$ begin with the word 
$a_0 \ulcorner\phi\urcorner a_0$.

1) An occurrence of word $xb$ where $x\notin\{ b,\b\}$ originally arises 
in the output
of Group 1: $\ a_0 \ulcorner\phi\urcorner a_0$.

2) The first active command from Group 2 operating in the evolution 
in question is $(e,b,?)\ar\tilde\nu$ , so the input of Group 2 has the form
$a_0 \ulcorner\phi\urcorner a_0$.
\nn

Let $\phi$ contain Boolean variables $\a _1 ,\a _2 ,\dots ,\a _q$.
\n

\proclaim{Lemma 2} There exists CA $\Cal A _2$ such that 
1) For any active command of $\Cal A _2 : \ (x,y,z)\ar w \ \ y,w\notin\w _1$ 
and if $t\in\{ x,z\} ,t\in\w _1$, then $t=\%$.
2) For any input data of the form (4) from the statement of Lemma 1 the 
following 
conditions are equivalent:

A). There exists a complete evolution of $\Cal A _2$ of the form
$$
B_j =B'_j ,B'_{j+1} ,\dots ,B'_{N'} ,
\tag 5$$

 B). $B_j$ contains an occurrence of the word $S\% \phi C$, where the end of 
the length $q$ of the word $S$ constitutes the list of values for Boolean 
variables which makes $\phi$ true.
3) Any complete evolution of $\Cal A _2$ of the form (5) has the length 
$N'=O(|\ulcorner\phi\urcorner |^2 )$.
\endproclaim
\nn

\head Proof of Lemma 2 \endhead
\nn

$\Cal A _2$ realizes a conventional procedure for the search of true value 
for $\phi$. Such a procedure is well known and it requires a quadratic time 
on Turing Machines.
 Lemma 2 is proved. 
\nn

Returning to the proof of inequality (2), note that if the list of active
commands for LSN $\Cal A$ is obtained from such lists for $\Cal A _1$ and 
$\Cal A _2$ 
by a simple join of them, then $\Cal A$ recognizes the satisfability of 
Boolean 
formulae in a quadratic time that implies inequality (2).

 It is evident that $\PLSN\subseteq\NPLSN$. So, the establishment of the 
following inclusion
$$
\NPLSN\subseteq\NP
\tag 6$$
would suffice to accomplish the proof of Theorem 1. We shall take up this 
subject now.
\n

Given NLSN $\Cal A$ computing the set $M_\Cal A$ in a polynomial time, 
one need only to obtain a nondeterministic cellular automaton $\Cal A ^*$, 
which computes $M_\Cal A$ in a polynomial time.
\n

Let
$$
B_1 ,B_2 ,\dots ,B_N
\tag 7$$
be an evolution of NLSN $\Cal A$ and let $W_i$ be some selected and 
underlined occurrence in 
$B_i : \ B_i =B'_i \underline{A'_i a_{j_i} A''_i}B''_i$
for each $i=1,2,\dots ,N$, so that 

1) for all $i=1,\dots ,N$

if $ B'_i \neq \Lambda$, then $|A'_i |=N+1$,

if $ B''_i \neq \Lambda$, then $|A''_i |=N+1$, where $\Lambda$ denotes the 
empty word,

2) for all $i=1,\dots ,N-1\ W_{i+1}$ is contained in the descendant of $W_i$ 
in the transformation 
$B_i \ar B_{i+1}$,

3) if $U_i$ is separated occurrence of $a_{j_i}$ in $W_i$ then $U_{i+1}$ is 
contained in the descendant of $U_i$ in the transformation $B_i \ar B_{i+1}$.

In this case the sequence of occurrences
$$
W_1 ,W_2 ,\dots W_N
\tag 8$$
is called a local sequence for the evolution (7), and the sequence
$$
a_{j_i} ,a_{j_2} ,\dots a_{j_N}
\tag 9$$
is called the central sequence for evolution (7) corresponding to the local 
sequence (8).

The following property of NLSN follows immediately from the definitions.
\nn

\proclaim{Property of localization} 
Let the sequence (9) be the central sequence for evolution (7) corresponding 
to the local sequence (8). Then for any $i\in\{ 1,\dots ,N\}$ and for any 
words $C'_i ,C''_i$
the sequence $a_{j_i}, a_{j_{i+1}} ,\dots ,a_{j_N}$ will be a central sequence 
for evolution
$B'_i ,B'_{i+1} ,\dots ,B'_N$ , where $B'_i =C'_i A_i a_{j_i} A''_i C''_i$.
\endproclaim
\n

This property makes it possible to simulate NLSN $\Cal A$ on a 
nondeterministic cellular automaton
$\Cal A ^*$. The work of $\Cal A ^*$ consists of sequential cycles, each of 
them
has an input $I_i$ and an output $I_{i+1}$ with separated occurrences.

Let $I_i =a_0 A'_i \underline{a_{j_i}} A''_i a_0$ be the input of some
step number $i$ with the separated occurrence, and 
$a_0 B'\underline{A}B''a_0 \in\Cal A (I_i )$ where $\underline{A}$ is the 
descendent of $\underline{a_{j_i}}$. Let $A=A_1 a_{j_{i+1}} A_2$ be some 
representation of $A$. 

Let $t_\Cal A (n)=O(n^p ) ,\ p\in\Bbb N ,\ M=|I_1 |$, where $I_1$ is an 
input of $\Cal A$.

Let $A'_{i+1}$ be the maximal from the ends of $B'A$, whose lengths do not 
exceed $t_\Cal A (M)+1$.

Let $A''_{i+1}$ be the maximal from the beginnings of $A_2 B''$, whose lengths 
do not exceed $t_\Cal A (M)+1$. Then
$$
I_{i+1} =a_0 A'_{i+1} \underline{a_{j_{i+1}}} A''_{i+1} a_0 .
$$

The automaton $\Cal A ^*$ described computes the set $M_\Cal A$ due to 
Property of localization.
It is readily seen that $\Cal A ^*$ requires the time $O(n^{2p} )$. Theorem 1 
is proved.

Note that Theorem 1 gives $\PLS\subseteq EXPTIME$ in particular.
\nnn

\specialhead \ \ \ \ \ \ \ \ \ \ 3. Nondeterministic version of LSN 
\endspecialhead
\nn

\subsubhead Theorem 2: Sketch of the proof \endsubsubhead
\n

A conventional procedure of diagonalization will be run by NLS.
Namely, if $\ulcorner\Cal B\urcorner$ denotes a code of CA $\Cal B$,
we shall construct such NLS $\Cal A$ with the time complexity
$t_\Cal A (n)=O(n)$ that for any alphabet $\w '$ and 
for any $\a =1,2,\dots$ there exist
$c_1 ,c_2 >0$ so that for any CA $\Cal B$ in alphabet $\w '$ the following 
two conditions are equivalent. 

1). There exists a complete evolution of $\Cal B$ of the form
$$
B'_0 =a_0 \ulcorner\Cal B\urcorner a_0 ,B'_1 ,\dots ,B'_M
$$
where $m=|\ulcorner\Cal B\urcorner | , M\leq exp(c_1 m)$, 
$\forall i=1,\dots ,M$
$|B'_M | \leq m^\a$ and $B_M$ contains
$a_{n-1}$.

2). There exists a complete evolution of $\Cal A$ of the form
$$
B_0 =a_0 \ulcorner\Cal B\urcorner a_0 ,B_1 ,\dots ,B_h ,\ h\leq c_2 m^\a .
$$

\n
I drop all details.
\nn

\specialhead \ \ \ \ \ \ \ \ \ \ 4. LS and the polynomial hierarchy 
\endspecialhead
\nn

There is an intimate connection between LS and the well-known polynomial 
hierarchy (PH).

PH is determined by the sequences of classes:
$$
\aligned \SP _0 &\subseteq\SP _1 \subseteq\dots ,\\
\PP _0 &\subseteq \PP _1 \subseteq \dots ,\endaligned
$$
defined by the following induction.

{\it Basis.} $\SP _0 =\PP _0 =\P$ - is the class of all predicates, computed
in a polynomial time on Turing Machines.

{\it Step.} 1. Let $\SP _n$ be the class of all predicates 
$A(x_1 ,\dots ,x_s )$
such that
$$
A(x_1 ,\dots ,x_s )\Longleftrightarrow\exists x_{s+1} :\ |x_{s+1} 
|\leq p(|x_1 |,\dots |x_s |)\ \
B(x_1 ,\dots ,x_s ,x_{s+1} )
$$
for some polynomial $p$ and some $B\in\PP _{n-1}$.

2. Let $\PP _n$ be the class of all predicates $A(x_1 ,\dots ,x_s )$
such that
$$
A(x_1 ,\dots ,x_s )\Longleftrightarrow\forall x_{s+1} :\ |x_{s+1} 
|\leq p(|x_1 |,\dots |x_s |)\ \
B(x_1 ,\dots ,x_s ,x_{s+1} )
$$
for some polynomial $p$ and some $B\in\SP _{n-1}$.

\n
Then $\PH=\bigcup\limits _{n=0}^{\infty}\SP _n =
\bigcup\limits _{n=0}^{\infty} \PP _n$.

Adding the sign of coma to alphabet $\w$ we naturally extend the class PLS 
of predicates to predicates $A(x_1 , \dots ,x_n )$ on 
$\underbrace{\w ^* \times\dots\times\w ^*}_{ n\ \text{times}}$ for any 
$n=1,2,\dots .$
\n

\proclaim{Theorem 3} $\PH\subseteq\PLS$.\endproclaim

\nn
\head Proof\endhead
\nn

It is sufficient to prove that for all $n=0,1,\dots$ the following two 
inclusions take place:
$$
\SP _n \subseteq\PLS
\tag 10
$$
$$
\PP _n \subseteq\PLS .
\tag 11
$$

This will be proved by a simultaneous induction on $n$.

\n
{\bf Basis} Follows from Theorem 1.

{\bf Step}

1) Let us prove the inclusion (10).

Given LS $\Cal B$ computing a predicate $B(x_1 ,\dots ,x_s ,x_{s+1}$ in a 
polynomial time
$p_1$ by the inductive hypothesis, the LS $\Cal A$ must be constructed
computing in a polynomial time the predicate
$$
A(x_1 ,\dots ,x_s )\Longleftrightarrow\forall x_{s+1} :\ |x_{s+1} 
|\leq p(|x_1 |,\dots |x_s |).
$$

We can suppose that $p$ and $p_1$ are increasing positive functions.

It will be readily seen how to define such  LS $\Cal A$ with the help of 
new auxiliary
letters in the following sequential steps.
\nn

{\bf Step 1}  {\it Input} : 
$$
x_1 ,x_2 ,\dots ,x_s
$$

{\it Output} 
$$
\a\tilde x_1 ,\dots ,\tilde x_s ,\tilde x_{s+1} \% \tilde x_1 ,\dots ,
\tilde x_s ,\tilde x_{s+2} \%
\dots \% \tilde x_1 ,\dots ,\tilde x_s ,\tilde x_{s+M} \a ,
\tag 12 $$
where $M\leq n^{p(|x_1 | ,\dots ,|x_s | )+1} +1,$ $x_{s+1} ,\dots ,x_M$ 
are all
the different words of the length $\leq p(|x_1 |,\dots ,|x_s |)$,
$x_i =y_{i1} y_{i2} \dots y_{is_i} ,$ all $y_{ij}$ are letters,
$\tilde x_i =Ty_{i1} Ty_{i2} T\dots Ty_{is_i} T,$ 
$ T=\underbrace{\b\times\dots\times\b}_{p_1}$,
$p_2 =p_1 (|x_1 |,\dots ,|x_s |,p(|x_1 |,\dots ,|x_s |)).$ $T$ will be a 
counter
recognizing the finish of the work of $\Cal B$ on the list 
$x_1 ,\dots x_{s+j} ;\ j=1,\dots ,M$.

Appropriate rules for $\Cal A$ can be written as in section 2.

\n
{\bf Step 2} {\it Input} :
(12)

{\it Output}:
$$
\a R_1 \% R_2 \%\dots \% R_{s+M} \a ,
\tag 13 $$
where $R_i$ is the result of the work of $\Cal B$ on $x_1 ,\dots , x_{s+j} ,\
 j=1,\dots ,M$. The counters are contracted on 1 in any substitution of 
$\Cal A$ in Step 2.
\n

{\bf Step 3} 

{\it Input} : (13)

{\it Output} contains the letter $\g$ iff $A(x_1 ,\dots x_s )$ is right.

Let $a_n$ be the final letter for $\Cal B$ if success. Rules for the Step 3 
have the form:
$$
\aligned
(x,y,z)&\ar\L ,\ \text{if} \ \% ,a_n ,\a \notin\{ x,y,z\} ,\ \text{or} \ 
y=z=a_n ,\\
(x,y,z)&\ar a_n ,\ \text{if only one from} \ x,y,z \ \text{is}\ a_n ,\\
(\% ,a_n ,\% )&\ar\L ,\\
(a_n ,\% ,a_n )&\ar\L ,\\
(a_0 ,\a ,\a )&\ar \g ,\\
(\a ,\% ,a_n )&\ar \L ,\\
(a_n ,\% ,\a )&\ar \L .
\endaligned
$$

Inclusion (10) is proved.
\n

Let's prove (11). 

The difference from the previous case is only in Step 3. Rules have the form:

$$
\aligned
(x,y,z)&\ar\L ,\ \text{if} \ a_n \notin\{ x,y,z\} ,\ \text{or}\ y=z=a_n ,\\
\text{in other cases}\ (x,y,z)&\ar a_n ,\ \text{if} \ a_n \in\{x,y,z\} ,\\
(\a ,a_n ,\a )&\ar \g .
\endaligned
$$

Theorem 3 is proved.
\nnn

\specialhead \ \ \ \ \ \ \ \ \ \ \ 5. Resume \endspecialhead
\nn

Thus, writing out sequentially the standard classes P, NP, HP and all the 
complexity
classes defined above, we obtain the following chain
$$
\P \overset ?\to\subseteq \NP =\PLSN =
\NPLSN\overset ?\to\subseteq\PH\overset ?
\to\subseteq\PLS\overset ?\to\subseteq\NPLS \neq\P _\a .
$$
The following two problems: $\P \overset ?\to\subseteq \NP$ and $\NP \overset
 ?\to\subseteq \PH$ are the famous open problems in the theory of complexity.
Therefore, in view of Theorem 3, Lindenmayer system (with the 
possibility of empty
results) may be even more powerful tool of computations than nondeterministic
Turing Machine.

Note that the nondeterminism of LS is conceptually classical. 
Harnessing of quantum mechanical effects for increasing the power of 
LS faces problems. The point is that reproductions of quantum states is 
forbidden by uncertainty principle.
\nn

\specialhead \ \ \ \ \ \ \ \ \ \ \ 7. References \endspecialhead

\n

\enddocument